\documentclass[conference]{IEEEtran}
\IEEEoverridecommandlockouts
\pdfoutput=1
\usepackage{cite}
\usepackage{amsmath,amssymb,amsfonts}
\usepackage{algorithmic}
\usepackage{graphicx}
\usepackage{textcomp}
\usepackage{xcolor}
\def\BibTeX{{\rm B\kern-.05em{\sc i\kern-.025em b}\kern-.08em
    T\kern-.1667em\lower.7ex\hbox{E}\kern-.125emX}}
\definecolor{ao(english)}{rgb}{0.0, 0.5, 0.0}

\usepackage{tabularx}
\usepackage{mathtools, nccmath}

\usepackage{multirow}
\usepackage{bm}
\usepackage{xspace}
\def\framework{PristiQ\xspace} % {QuantumFlow\xspace}
\def\pricirc{\textit{PriCircuit}\xspace}
\def\pricomp{\textit{PriCompiler}\xspace}
\def\primod{\textit{PriModel}\xspace}
\def\CNOT{CNOT\xspace}
    
\begin{document}

\title{PristiQ: A Co-Design Framework for Preserving Data Security of Quantum Learning in the Cloud
\thanks{Pristiq is a prescription medicine to treat depression. We envision our framework to treat the depression by data leakage in future quantum machine learning research.}
}

\author{\IEEEauthorblockN{
Zhepeng Wang\textsuperscript{\dag, \S},
Yi Sheng\textsuperscript{\dag},
Nirajan Koirala\textsuperscript{\ddag}, 
Kanad Basu\textsuperscript{*}, 
Taeho Jung\textsuperscript{\ddag}, 
Cheng-Chang Lu\textsuperscript{\P}, 
Weiwen Jiang\textsuperscript{\dag, \S}
% Jinjun Xiong\textsuperscript{\P},
}

\IEEEauthorblockA{\textsuperscript{\dag}Department of Electrical and Computer Engineering, George Mason University\\
\textsuperscript{\S}Quantum Science and Engineering Center, George Mason University\\
\textsuperscript{\ddag}Department of Computer Science and Engineering, University of Notre Dame\\ 
\textsuperscript{*}Department of Electrical \& Computer Engineering, University of Texas at Dallas\\
\textsuperscript{\P}Qradle Inc.\\
\{zwang48, wjiang8\}@gmu.edu
\vspace{-0.15in}}
}

\maketitle

\begin{abstract}
Benefiting from cloud computing, today's early-stage quantum computers can be remotely accessed via the cloud services, known as Quantum-as-a-Service (QaaS). However, it poses a high risk of data leakage in quantum machine learning (QML). To run a QML model with QaaS, users need to locally compile their quantum circuits including the subcircuit of data encoding first and then send the compiled circuit to the QaaS provider for execution. If the QaaS provider is untrustworthy, the subcircuit to encode the raw data can be easily stolen. Therefore, we propose a co-design framework for \underline{Pr}eserv\underline{i}ng the data \underline{s}ecuri\underline{ty} of \underline{Q}ML with the QaaS paradigm, namely \framework. By introducing an encryption subcircuit with extra secure qubits associated with a user-defined security key, the security of data can be greatly enhanced. And an automatic search algorithm is proposed to optimize the model to maintain its performance on the encrypted quantum data. Experimental results on simulation and the actual IBM quantum computer both prove the ability of \framework to provide high security for the quantum data while maintaining the model performance in QML.

\end{abstract}

\begin{IEEEkeywords}
Cloud Quantum Computing, Quantum Machine Learning, Quantum Data Security
\end{IEEEkeywords}
\vspace{0mm}
\section{Introduction}\label{Sec:Introduction}
\begin{figure*}[!t]
\centering
\includegraphics[width=6.8in]{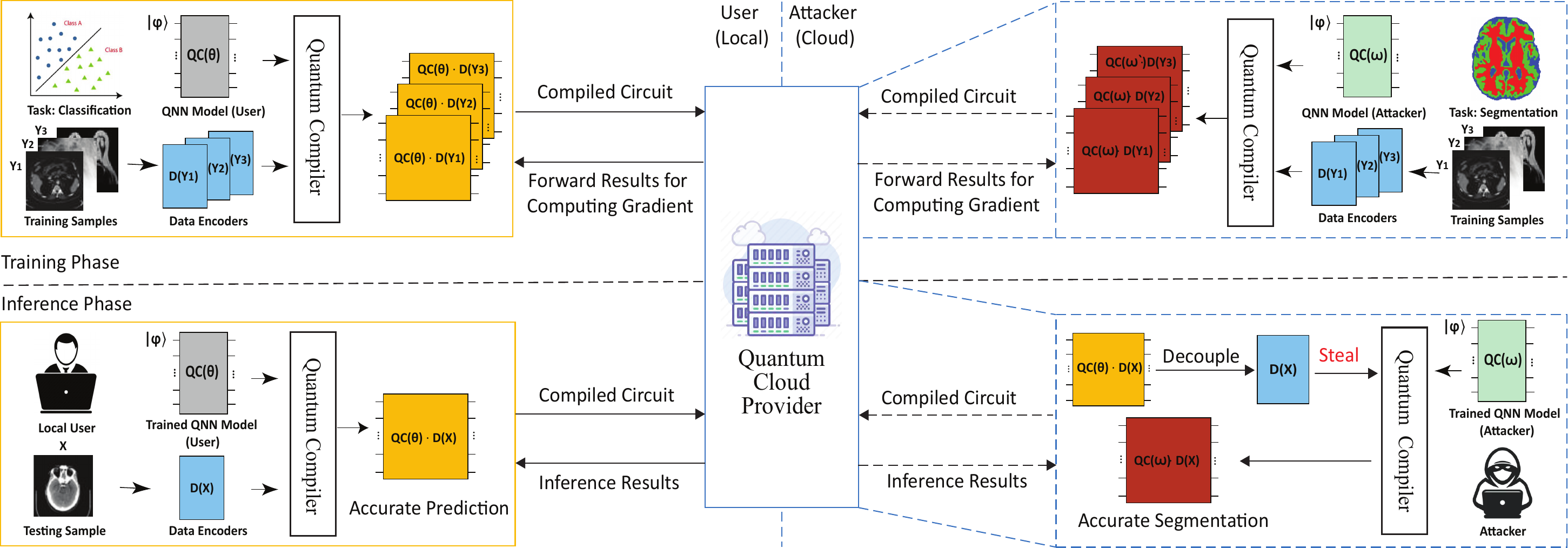}
\caption{Illustration of the typical threat model in QaaS paradigm for QML applications: The attacker in the untrustworthy cloud quantum provider can steal the user data for its own QML task. }
\label{fig:qprivate}
\end{figure*}
Over decades of development in quantum science, we are now witnessing the breakthrough of technologies in designing quantum computers: the number of qubits in the actual quantum computers has increased rapidly, from 5 qubits in 2016 to 433 qubits in 2022~\cite{ibm433q}. With access to these actual quantum computers via the cloud services, known as Quantum-as-a-Service (QaaS), various studies have been done to explore the quantum advantage in different problems~\cite{shor1999polynomial,zhong2020quantum, Zhan_2024}, where quantum machine learning (QML), in particular, quantum neural network (QNN) has attracted great research efforts recently~\cite{sim2019expressibility,jiang2021co, zhan2023quantum, hu2022design, li2023novel, wang2022quest}.

Although QNN is promising in many real applications, the current QaaS paradigm for the execution of QNN poses a high risk of data leakage due to the raw data delivery to the cloud. Fig.~\ref{fig:qprivate} shows a typical threat model in the QaaS paradigm for QML applications. In this case, the local user and the attacker may train their own QNN models on the same data but target different tasks. For example, with the same medical images, the user could build a QNN for classification while the attacker could build a different QNN for segmentation. The training process of a QNN is usually executed in a hybrid quantum-classical way. For instance, the classical data $\bm{Y_1}$ will first be encoded as a quantum circuit $D(\bm{Y_1})$ with a data encoding method (e.g., amplitude encoding). $D(\bm{Y_1})$ and the quantum circuit $QC(\bm{\theta})$ 
%(or $QC(\omega)$) 
for the QNN will be provided to the local quantum compiler, which will generate a compiled quantum circuit $QC(\bm{\theta}) \cdot D(\bm{Y_1})$. 
%(or $QC(\omega) \cdot D(Y_1)$)
The compiled circuit is then sent to the quantum cloud provider to run the forward pass. And the forward results will be sent back to the user to compute the gradients of learnable parameters $\bm{\theta}$ through parameter shift~\cite{mitarai2018quantum}. For the quantum circuit  $QC(\bm{\omega})$ of the attacker, it can be trained in the same way. 

After the training process, the compiled circuit $QC(\bm{\theta}) \cdot D(\bm{X})$ will be sent to the quantum cloud provider for inference to get an accurate classification of sensitive user data $\bm{X}$. However, if the quantum cloud provider is untrustworthy, the attacker can directly get access to $QC(\bm{\theta}) \cdot D(\bm{X})$. The encoding quantum circuit $D(\bm{X})$ can be decoupled from $QC(\bm{\theta}) \cdot D(\bm{X})$ since the data encoding algorithms have fixed circuit patterns that are easy to be detected. Moreover, with $QC(\bm{\omega})$, an accurate segmentation of $D(\bm{X}$ can be made, which means that the attacker can obtain more information from $D(\bm{X})$ with data analysis methods. Such kind of data leakage can not only cause privacy breaches to the users but also bring huge damage to the reputation of the institution that leverages the power of QNNs for their applications. Therefore, the preservation of data security is of great importance in QML applications.

There exist many works on blind quantum computing~\cite{ fitzsimons2017private,broadbent2009universal}, which focus on the security of quantum circuits. However, they assume that the user has a lightweight quantum terminal to preprocess the data in quantum states, which does not hold in the current QaaS paradigm for quantum applications. Traditional secure computation methods like homomorphic encryption~\cite{gilad2016cryptonets, peng2024lingcn} cannot be directly applied to quantum computing since the involved complicated computing operations cannot be implemented by quantum operations. Therefore, a quantum computing-based methodology is necessary to protect the raw input data to the QNNs in the context of QaaS. 

In this paper, we make the very first attempt to preserve data security in QML applications by formulating a circuit-compiler-model cross-layer design framework, namely \framework. The proposed framework is composed of three components: \textit{i)} \pricirc incorporates an encryption subcircuit into the subcircuit of data encoding to keep data security; \textit{ii)} \pricomp obfuscates the encryption subcircuit; and \textit{iii)} \primod automatically searches for the optimal design of the QNN model in order to maintain the performance when directly conducting computations on the encrypted data. These components will work together to provide data security while maintaining high performance for the QML tasks through QaaS.

% ~\todo{I delete the summary of experimental results here like QuantumFlow}

% \textbf{\textit{Results:}}

% \todo{}

% PristiQ is evaluated on MNIST and Fashion datasets.
% First of all, by using \textit{PriCircuit} and \textit{PriCompiler} only, we can achieve up to 1.8$\times$ average reduction on the peak signal-to-noise ratio (PSNR) for data encryption; while, on MNIST dataset with 3 sub-classes (i.e., MNIST-3), it suffers up to 4.82$\%$ accuracy loss from 93.82$\%$ to 89.00$\%$ due to the data encryption.
% Then, by integrating \textit{PriModel} into PristiQ, the QNN accuracy can even be improved from 93.82\% to 95.04\%.

% \todo{Highlight it separately?}
% \textbf{\textit{Contributions:}} 

The main contributions of this paper are as follows.
\begin{itemize}
    \item The first and the main contribution of this paper is the proposal of \framework, which is, to the best of our knowledge, the first framework to secure input data to QNNs under QaaS paradigm.
    \item \framework provides a fundamental understanding of the design of a secure quantum computing system: a cross-layer co-design is required to preserve data security in quantum computing. This design philosophy is not limited to QML applications but can also be applied to other types of quantum applications.
    % \item Most of the existing quantum compilers focus on the reduction of circuit length or the improvement of fidelity; \framework involves compiler optimization to support the data security, which brings a new potential optimization direction in the future design of the quantum compilers.
    \item \framework brings the concept of model adaptation to the design of QNN, which preserves high performance on the encrypted data.
\end{itemize}
% \vspace{-2pt}

% The rest of this paper is organized as follows. Section~\ref{Sec:Related Work} outlines the related works of this paper.  Then, in Section~\ref{Sec:Method}, the technical details of \framework are illustrated. Section~\ref{Sec:Experiment} shows the experimental results to evaluate \framework. And Section~\ref{Sec:Conclusions} remarks on the conclusion.

%-----------------------------------------

% The rest of this paper is organized as follows. First, in Section~\ref{Sec:Threat Model}, the threat model of executing QNN through QaaS is illustrated. The preliminary background and motivational examples are shown in Section~\ref{Sec:Preliminary}. Then, Section~\ref{Sec:Method} introduces the proposed PristiQ framework. Evaluation of PristiQ is conducted in Section~\ref{Sec:Experiment}. Section~\ref{Sec:Related Work} outlines the related works of this paper, and Section~\ref{Sec:Conclusions} remarks the conclusion.~\todo{We might need to adapt this part, especially Section 3}

% The rest of this paper is organized as follows. Section~\ref{Sec:Related Work} outlines the related works of this paper. Then, in Section~\ref{Sec:Threat Model}, the threat model of executing QNN through QaaS is illustrated. The preliminary background of quantum computing is shown in Section~\ref{Sec:Preliminary}. Section~\ref{Sec:Method} introduces the proposed PristiQ framework. Evaluation of PristiQ is conducted in Section~\ref{Sec:Experiment}. And Section~\ref{Sec:Conclusions} remarks the conclusion.
\section{Related Work}\label{Sec:Related Work}
\noindent\textbf{Quantum Neural Network.} Since the fundamental operation in deep learning on classical computing is neuron computation, \cite{cao2017quantum} proposed the very first quantum neuron to mimic the behavior of the classical neuron. Based on this work,~\cite{bausch2020recurrent} proposed the quantum recurrent neural network (QRNN), while QuantumFlow~\cite{jiang2021co} proposed a co-design framework to implement the quantum neuron onto a quantum processor with the demonstration of quantum advantage. There is another type of QNN, which has no classical counterparts, known as variational quantum circuits (VQCs)~\cite{sim2019expressibility, schuld2021effect, li2024quapprox}. The QNNs built with VQCs include 
the quantum generative adversarial networks (QGAN)\cite{dallaire2018quantum} and the quantum convolutional neural networks (QCNN)\cite{cong2019quantum}. These works provided the fundamentals and motivation for the research of QNNs.~\cite{wang2021exploration} proposed a hybrid quantum neuron architecture design by mixing these two types of quantum neuron designs to achieve better performance. The potential of QNNs has inspired many works to study it from different aspects. To make the QNNs available in the noisy intermediate-scale quantum (NISQ) era, ~\cite{liang2021can} conducted research on the robustness of Quantumflow on NISQ devices.~\cite{hu2022quantum} proposed a novel QNN compression algorithm to improve the performance of QNNs under quantum noise. The high variance of quantum noise across devices and time brought new challenges to the development of QML. Therefore, ~\cite{senapati2023towards} studied the reproducibility of QML while~\cite{hu2023toward} proposed calibration~\cite{hu2023toward} method.~\cite{ruan2023violet, ruan2023venus} are works on the visualization of quantum states and quantum noise, providing tools to further analyze the behavior of QML models with the temporal and spatial variance on the quantum noise. Moreover, although most of the existing works mainly focus on deploying QNNs for inference on quantum devices, there exist recent works on enabling the training of QNNs on quantum devices~\cite{wang2022qoc, wang2022quantumnat}.

\noindent\textbf{Secure Quantum Computing}. Techniques in blind quantum computing~\cite{fitzsimons2017private,broadbent2009universal} are proposed to preserve the security of quantum circuits under the assumption that the client has a local quantum terminal to rotate and measure a single qubit while the communication between the client and server relies on a quantum communication channel. However, these assumptions do not hold in the QaaS paradigm for QML. \cite{huang2017experimental} customizes a distributed method to protect the quantum circuit for prime factorization with classical local terminal only. But it cannot be directly extended to QML and it requires the involvement of multiple distributed quantum computers. \cite{mahadev2020classical} proposes an encryption algorithm to protect the data encoded by basis encoding, which is not an efficient and practical encoding method in QML.~\cite{wang2023qumos} proposed a distributed algorithm to preserve the security of the QML model in the cloud. But it focuses on the protection of the QML circuit instead of the quantum data circuit.

\noindent\textbf{Privacy-Preserving Machine Learning without Encryption}. Various approaches have been proposed to protect data privacy in the scenarios where machine learning is performed over distributed datasets possessed by users. Differential privacy \cite{abadi2016deep,arachchige2019local} ensures individual data is hidden from an untrusted aggregator by the users who add accuracy-dependent perturbation to the shared data. Federated learning \cite{yang2019federated,jiang2022federated} is another option which lets users shares model parameters with an untrusted aggregator instead of their datasets. There also exist a hybrid approach that combines the differential privacy and federated learning \cite{wei2020federated,zhao2020local}. In this case, the computation itself is not secured, but the data is preprocessed such that it does not leak sensitive information. Such work is orthogonal to our work where the computation process is secured by encrypting the input data.

\noindent\textbf{Secure Machine Learning with Secure Computation}. Some approaches secure the machine learning process by performing computations on encrypted data. This way, outsourced computation/data is secure against the cloud. There exist approaches based on homomorphic encryption \cite{gilad2016cryptonets, peng2024lingcn}, secure multiparty computation~\cite{riazi2018chameleon, peng2023rrnet,luo2023aq2pnn, peng2023autorep}, the combination of the previous two \cite{juvekar2018gazelle}, and trusted execution environment technologies \cite{tramer2018slalom,hashemi2020darknight}. These are the most similar to our work in terms of what is secured in the machine learning process. However, these focus on classical computing environments, and such encryption methodologies can hardly be applicable in quantum computing scenarios.
\section{PristiQ Framework}\label{Sec:Method}
\begin{figure*}[!t]
\centering
\includegraphics[width=5in]{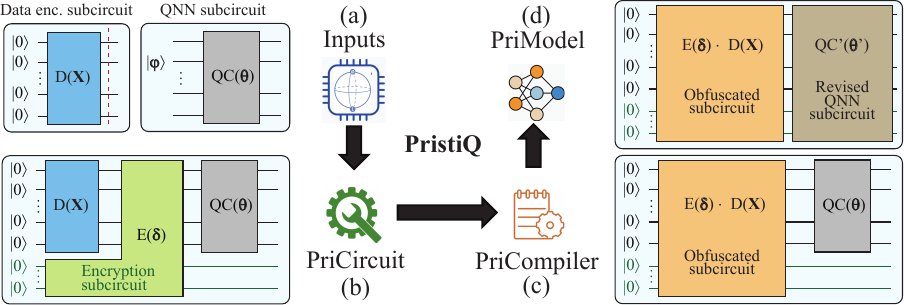}
\caption{Illustration of \framework framework: (a) the data encoding subcircuit $D(\bm{X})$ and QNN subcircuit $QC(\bm{\theta})$ as the inputs; (b) \pricirc, adding the encryption subcircuit $E(\bm{\delta})$; (c) \pricomp, obfuscating $D(\bm{X})$ and $E(\bm{\delta})$ by forming $E(\bm{\delta})\cdot D(\bm{X})$; (d) \primod, revising QNN subcircuit $QC(\bm{\theta})$ to $QC^{\prime}(\bm{\theta}^{\prime})$ for maintaining the performance on encrypted data.}
\label{fig:overview}
\end{figure*}

Fig.~\ref{fig:overview} shows an overview of \framework. Given the data encoding subcircuit $D(\bm{X})$ and QNN subcircuit $QC(\bm{\theta})$ as inputs, which is shown in Fig.~\ref{fig:overview} (a), \framework will sequentially go through the three components: \pricirc in Fig.~\ref{fig:overview}(b), \pricomp in Fig.~\ref{fig:overview}(c) and~\primod in Fig.~\ref{fig:overview}(d).

Based on $D(\bm{X})$ and $QC(\bm{\theta})$, \pricirc will reconstruct the original circuit $QC(\bm{\theta}) \cdot D(\bm{X})$ by inserting an encryption subcircuit $E(\bm{\delta})$ between $D(\bm{X})$ and $QC(\bm{\theta})$, which introduces extra qubits, as shown in Fig. \ref{fig:overview}(b). To avoid the attacker detecting the boundary between $D(\bm{X})$ and $E(\bm{\delta})$ easily, \pricomp is designed to obfuscate $D(\bm{X})$ and $E(\bm{\delta})$. More specifically, it utilizes the compiling optimization to form a merged circuit $E(\bm{\delta})\cdot D(\bm{X})$, making their boundary indistinguishable. In this way, the attacker could only extract the encrypted data encoded by $E(\bm{\delta})\cdot D(\bm{X})$ instead of the raw input data encoded by $D(\bm{X})$. However, directly applying the original QNN subcircuit $QC(\bm{\theta})$ to the encrypted data may incur significant performance degradation. Therefore, to maintain the performance on encrypted data, \primod will revise $QC(\bm{\theta})$ by replacing it with subcircuit $QC^{\prime}(\bm{\theta^{\prime}})$. The neural architecture of the QNN for $QC^{\prime}(\bm{\theta^{\prime}})$ is designed by using an automatic search engine based on reinforcement learning, aiming at the best performance on the encrypted data. In the following content of this section, we will introduce the details of all three components in~\framework.
\vspace{-3mm}

\subsection{PriCircuit}\label{Subsec:PriCircuit}
\noindent\textbf{Design Principle:} The main purpose of \pricirc is to provide data security, i.e., the raw input 
data should be hidden; however, the encrypted data should also preserve the critical information contained by the raw input data, such that the encrypted data can still be learned effectively.
Unlike conventional neural networks in classical computing, most quantum neural networks~\cite{jiang2021co,cao2017quantum} do not extract the spatial features. Therefore, the most important information in the raw input data is the relative relationship among its features.
Based on this observation, we propose a two-stage encryption process for \pricirc. And the details are as follows.

\begin{figure}[!t]
\centering
\includegraphics[width=3.3in]{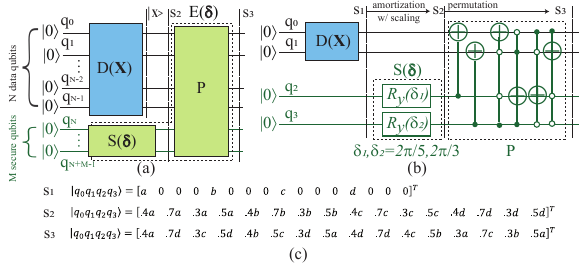}
\caption{\pricirc: (a) encryption subcircuit $E(\bm{\delta})$ composed of $S(\bm{\delta})$ and $P$; (b) an example of \pricirc with 2 data qubits and 2 secure qubits; (c) transformations of quantum states: $S_1\rightarrow S_2$ and $S_2\rightarrow S_3$.}
% Illustration of threat model: An adversary obtains the computing quantum circuit and reconstruct the data (e.g., MNIST) by creating a reverse circuit. }
\label{fig:circuit}
\end{figure}

\noindent\textbf{Design overview:} Fig.~\ref{fig:circuit} shows the detailed design of encryption subcircuit $E(\bm{\delta})$ generated by \pricirc with an example.
As shown in Fig.~\ref{fig:circuit}(a), there are $N$ qubits used to encode raw input data $\bm{X}$ with $2^N$ features, named data qubits. And $M$ extra qubits are added to the $N$ data qubits, called secure qubits.
The first stage applies the subcircuit $S(\bm{\delta})$, which is a part of $E(\bm{\delta})$,  to operate on the newly added $M$ secure qubits with random parameter set $\bm{\delta}$ to perform amortization on the amplitudes from data qubits to secure qubits; then, the second stage utilizes the remaining part of $E(\bm{\delta})$ (i.e., subcircuit $P$) to perform permutation on the amplitudes of the quantum state processed by $S(\bm{\delta})$ since the permutation matrix can be implemented by quantum circuits. Based on the design of \pricirc, we define the security key in \framework as a tuple  $\langle \bm{\delta}, P\rangle$, where $\bm{\delta}$ is the random parameter set used in the amortization with scaling while $P$ is the design of random permutation circuit.

\textbf{Stage 1: amortization w/ scaling.} As shown in Fig.~\ref{fig:circuit}(a), stage 1 starts with initial quantum states $S_1$, where $S_1 = |\bm{X}\rangle \otimes |0\rangle^{\otimes M} = |q_0q_1 ... q_{N-1}q_{N} ... q_{N+M-1} \rangle$. Here $|\bm{X}\rangle$ is the encoded quantum state for the raw input data $\bm{X}$, using the data encoding subcircuit $D(\bm{X})$. Note that for $S_1$, only quantum basis state $|q_0q_1 ... q_{N-1}0 ... 0\rangle$ has non-zero amplitudes since no quantum gates are applied to the $M$ secure qubits. For example, in Fig.~\ref{fig:circuit}(b), the number of data qubits $N$ is 2 while the number of secure qubits $M$ is 2 as well. In this case, $S_1 = |q_0q_1q_2q_3 \rangle = |\bm{X}\rangle \otimes |00\rangle $ and only quantum state $|q_0q_100\rangle$ has non-zero amplitudes, which is clearly shown in the state vector of $S_1$ in Fig.~\ref{fig:circuit}(c).

Stage 1 focuses on the transformation from state $S_1$ to state $S_2$, where each secure qubit is operated by $Ry$ gate with a parameter that is randomly generated. More specifically, the $k^{th}$ secure qubit is denoted as $Sec_{k}$ and its initial state is $|0\rangle$. After applying an $Ry$ gate with a parameter $\delta_{k}$ to it, the state of $Sec_{k}$ is converted to ${\lbrack \cos \frac{\delta_{k}}{2}, \sin \frac{\delta_{k}}{2} \rbrack}^{T}$, which is denoted by $|QS_{k}\rangle$. Therefore, the quantum state $|Q\rangle$, which is the state composed of all the secure qubits, can be computed as,
\vspace{-3pt}
\begin{equation}
    \begin{split}
    |Q\rangle &= |QS_{0}\rangle \otimes |QS_{1}\rangle \otimes ...\otimes |QS_{k}\rangle \otimes ...\otimes |QS_{M-1}\rangle . 
    \end{split}~\label{eq:pric_Q}
\end{equation}
\vspace{-3pt}
By combining the quantum states of data qubits $|\bm{X}\rangle$ and secure qubits $|Q\rangle$, we have $S_2 = |\bm{X}\rangle \otimes |Q\rangle$, where $|Q\rangle$ is served as the \textbf{scaling} vector with $2^M$ scaling factors for the \textbf{amortization} on the amplitudes from data qubits to secure qubits. For example, in Fig.~\ref{fig:circuit}(b), the amplitude on $|q_0q_100\rangle$ is amortized to the other three states (i.e., $|q_0q_101\rangle$, $|q_0q_110\rangle$, $|q_0q_111\rangle$) with scaling factors in $|Q\rangle$ determined by the parameter set $\bm{\delta}$. For instance, in Fig.~\ref{fig:circuit}(b), $|\bm{X}\rangle=[a,b,c,d]^{T}$. Since $\bm{\delta} = [\delta_0, \delta_1] = [\frac{\pi}{5}, \frac{\pi}{3}]$, we have $|QS_0\rangle=[cos\frac{\pi}{5},sin\frac{\pi}{5}]=[0.81,0.59]$ and $|QS_1\rangle=[cos\frac{\pi}{3},sin\frac{\pi}{3}]=[0.5,0.87]$. According to Equation~(\ref{eq:pric_Q}), we have $|Q\rangle=[0.4,0.7,0.3,0.5]$. We can then obtain $S_2=[a\cdot|Q\rangle,b\cdot|Q\rangle,c\cdot|Q\rangle,d\cdot|Q\rangle]^{T}$. The specific value of $S_2$ is shown in Fig.~\ref{fig:circuit}(c). And there exist four groups of values in $S_2$, where the values in each group are multiplied with a unique amplitude from $|\bm{X}\rangle$. Since all the groups share the same scaling vector $|Q\rangle$, the relative magnitude between arbitrary pair of groups is equal to that between the unique amplitudes they multiply with. In this way, the relative relationship between features within raw input data $|\bm{X}\rangle$ is maintained. 

% \begin{equation}
%     \begin{split}
%      S_2 &= |\bm{X}\rangle \otimes |Q\rangle \\
%      \end{split}~\label{eq:pric_s2},
% \end{equation}

% $|Q\rangle =|QS_0\rangle\otimes |QS_1\rangle=[0.4,0.7,0.3,0.5]$

% At state $s_2$, both data qubits and encryption qubits have amplitudes. We can compute the state of the circuit by using the scaling vector $|Q\rangle$.
% More specifically, after the data encoding circuit $D(\bm{X})$, we have the state of data qubits to be $|\bm{X}\rangle$.
% Denote the state of the full circuit at $S_2$ to be $|psi\rangle$, we can compute $|psi_1\rangle$ as follows.
% \begin{equation}
%     \begin{split}
%      |\psi_1\rangle &= |\bm{X}\rangle \otimes |Q\rangle \\
%      \end{split}
% \end{equation}

\textbf{Stage 2: permutation.} Although the random parameter set $\bm{\delta}$ is unknown to the attacker, stage 1 is not sufficient for the encryption of data since the attacker could get % the scaling vector $|Q\rangle$ or the random parameter set 
$\bm{\delta}$ and $|\bm{X}\rangle$ by solving a system of trigonometric equations if they know the processing of stage 1 is applied. Therefore, \pricirc further generates quantum state $S_3$ by randomly permuting the amplitudes in $S_2$ with a permutation matrix $P$. Since permutation matrix $P$ is unknown to the attacker, the system of trigonometric equations cannot be formulated and thus not be solved. Note that because the permutation matrix is a unitary matrix, it is always feasible to implement it with a corresponding quantum circuit, as shown in Fig.~\ref{fig:circuit}(a). Therefore, we have $S_3 = P \cdot S_2.$. For the example in Fig.~\ref{fig:circuit}(b), it utilizes multiple \CNOT gates to implement a permutation matrix and the specific value of $S_3$ after permutation is shown in Fig.~\ref{fig:circuit}(c). And it is obvious that there is no explicit group of values that shares the same multiplier after the permutation.

\vspace{-3mm}
\subsection{PriCompiler}\label{Subsec:PriCompiler}
\begin{figure}[!t]
\centering
\includegraphics[width=3.3in]{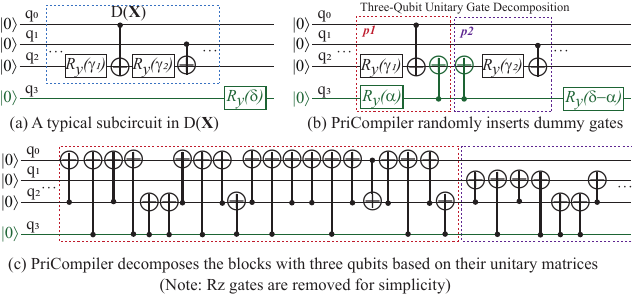}
\caption{\pricomp: (a) a typical subcircuit in $D(\bm{X})$; (b) the subcircuit with dummy \CNOT gates; (c) the compiled subcircuit of (b) using \pricomp.
}
\label{fig:compiler}
\end{figure}
\noindent\textbf{Potential Threat and Design Principle:} The main purpose of \pricomp is to obfuscate $E(\bm{\delta})$.
This is motivated by the fact that the boundary between the secure qubits and data qubits might be detected based on the design of \pricirc since there are only $Ry$ gates on the secure qubits in $S(\bm{\delta})$ while the data qubits on $D(\bm{X})$ contains multiple two-qubit gates. If the boundary is identified, then $D(\bm{X})$ can be extracted.

Therefore, we propose to introduce two-qubit gates on the secure qubits in $S(\bm{\delta})$, making the boundary between secure qubits and data qubits indistinguishable. However, naively adding more two-qubit gates may introduce extra undesired operations on the input data and thus disrupt the features for inference. Therefore, \pricomp should try to avoid actual interaction with $D(\bm{X})$, which brings new challenges for obfuscation. Fortunately, by reviewing the QaaS paradigm in Fig.~\ref{fig:qprivate}, we observe that the quantum compiler can be leveraged to make the obfuscation.

\noindent\textbf{Design Details:} \pricomp follows a two-stage design flow: (1) a dummy gate insertion and (2) unitary gate decomposition. Based on the fact that two consecutive \CNOT gates can be canceled out with each other (i.e., $I=CNOT\cdot CNOT$), we propose a method to create dummy gates and introduce two-qubit gates on the secure qubits. Fig.~\ref{fig:compiler} shows an example of the whole design flow.
The circuit in Fig.~\ref{fig:compiler}(a) is a typical subcircuit in $D(\bm{X})$. In this subcircuit, the three data qubits are operated with two-qubit gates while the single secure qubit only has an $Ry$ gate with a random parameter $\delta$.
In Fig.~\ref{fig:compiler}(b), an $Ry$ gate with two dummy \CNOT gates is added to the secure qubit for obfuscation. More specifically, we first add an $Ry$ gate with a randomly generated parameter $\alpha$. We then adjust the parameter of the original $Ry$ gate to $\delta-\alpha$ to ensure the functional correctness of the circuit. Between these two $Ry$ gates, we add two consecutive \CNOT gates. After this, the first stage is completed.

% to minimize the circuit length
In the second stage, given the circuit with dummy \CNOT gates from the first stage, the default quantum compiler will remove them. But in \pricomp, it adds a barrier between the pair of dummy \CNOT gates to split the circuit into two parts, i,e., $p_1$ and $p_2$, as shown in Fig.~\ref{fig:compiler}(b). In $p_1$ and $p_2$, each subcircuit will be assembled into a block if it includes exactly three qubits. In this example, $p_1$ and $p_2$ can be regarded as two blocks. Since all the gates in each block correspond to a unitary gate, the unitary matrix of each unitary gate can be calculated and further decomposed to a new circuit with different designs consisting of chosen basis gates. Following this rule, the resultant circuit compiled from the one in Fig.~\ref{fig:compiler}(b) is presented in Fig.~\ref{fig:compiler}(c).
In Fig.~\ref{fig:compiler}(c), it is clearly shown that multiple two-qubit gates are introduced on the secure qubit $q_3$. Due to the limited space, we only keep the \CNOT gate in the circuit in Fig.~\ref{fig:compiler}(c). And the consecutive \CNOT gates can thus not be canceled out due to the existence of $Rz$ gates between them.

\subsection{PriModel}
\begin{figure}[!t]
\centering
\includegraphics[width=3.3 in]{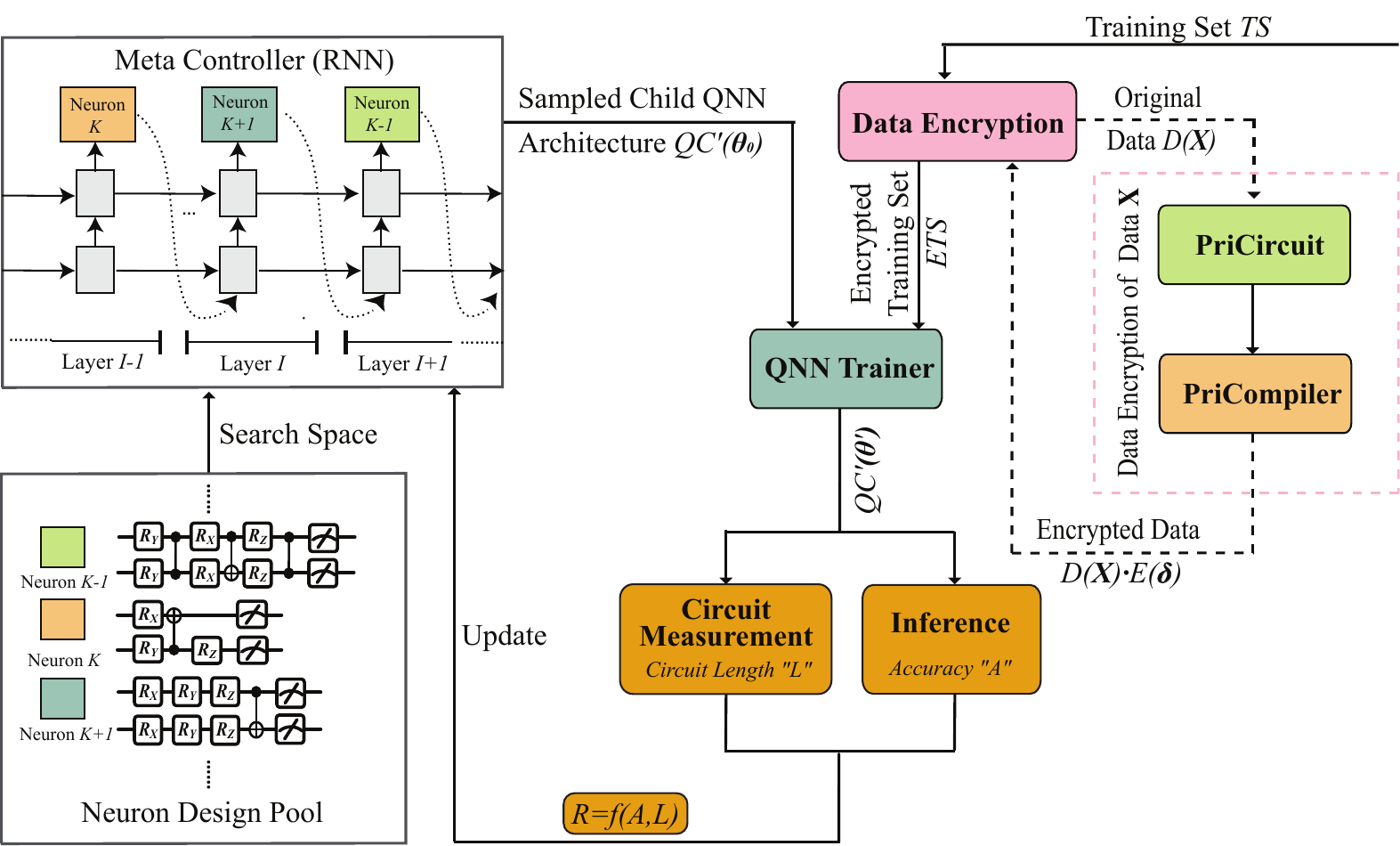}
\caption{The workflow of \primod}
\label{fig:primodel}
\end{figure}
After the process of \pricirc and \pricomp, the total number of input qubits is increased from $N$ to $N+M$. Therefore, the dimensions of features to the QNN model for processing is expanded from $2^N$ to $2^{N+M}$ as well. However, the original QNN subcircuit $QC(\bm{\theta})$ only contains operations on the $N$ data qubits without consideration of $M$ secure qubits, as shown in Fig.~\ref{fig:overview}(c). Moreover, the original QNN subcircuit $QC(\bm{\theta})$ is designed and optimized for the raw input data instead of the encrypted data. As a result, directly applying this QNN model to the encrypted data can degrade the performance of QNN model, which is shown in Section~\ref{Sec:Experiment}. Therefore, a reinforcement learning-based algorithm, namely \primod, is proposed to automatically search for the best QNN architecture for the encrypted data. 
% by \textit{PriCircuit} and \textit{PriCompiler}.

% This subsection will provide an optimization framework to identify the suitable model for the encrypted data with $N+M$ input qubits.~\todo{Justification of PriModel}

Fig.~\ref{fig:primodel} shows the overview of \primod. 
% to identify the best QNN architectures.
It is composed of three components: (1) a meta controller to guide the whole process which is implemented with a recurrent neural network (RNN), (2) a neuron design pool to serve as the search space for the architecture of QNN, which consists of multiple quantum neuron designs. (3) a security-aware QNN evaluator (i.e., the right part of Fig.~\ref{fig:primodel}) to measure each sampled solution from the meta controller.
% -based neural architecture search~\cnote{} to the design of QNNs and the overview of \textit{PriModel} is shown in . 
% In \textit{PriModel},

% \todo{do we need to claim data share the same encryption? more explain about inference on the encrypted data}

% \todo{More explanation about QNN training and the platform for accuracy of inference?}

More specifically, for the meta controller, the RNN cell in time step $I$ corresponds to the design of the quantum neuron in layer $I$ of the QNN, which generates a probability distribution for the sampling from the neuron design pool. The sampled neuron design for each layer will be connected to form a child QNN  $QC^{\prime}(\bm{\theta_{0}})$, which is initialized with $\bm{\theta_{0}}$ and sent to the evaluator. During the evaluation, $QC^{\prime}(\bm{\theta_{0}})$ will first be trained on the encrypted training set $ETS$, which is obtained by encrypting the data in the original training set $TS$. For each original data $D(\bm{X})$ in $TS$, it will go through \pricirc and \pricomp to get the encrypted data $D(\bm{X})\cdot E(\bm{\delta})$. And the outcome will be saved in $ETS$. The trained QNN model $QC^{\prime}(\bm{\theta^{\prime}})$ is then evaluated to calculate two metrics, i.e., circuit length $L$ and accuracy $A$. These two metrics are used for calculating the reward $R$ to update the meta controller. The reward function is defined as,

\begin{equation}
    R = A - b - \lambda\frac{L}{L_{base}}~\label{eq:reward},
\end{equation}
where $b$ is the exponential moving average of the accuracies of previous sampled QNNs. $L_{base}$ is a preset baseline circuit length for normalization. According to the available quantum computing resources, the user can use $\lambda$ to control the trade-off between the accuracy and the complexity of circuit implementation for the sampled QNN. 
% These steps will be repeated for multiple episodes until the meta controller converges or the preset maximum number of episodes is achieved. And the best QNN optimized for the encrypted data can thus be found after this process.

% \subsection{Put all together}
% \todo{Can we delete this part?}
% With the integration of \pricirc, \pricomp, and \primod into \framework, we can finally obtain the quantum circuit which preserves data security while achieving high accuracy on the target QML application. 

% The proposed framework will be evaluated in the next section.

%---------------------------------------------------------
% Topical subheadings are allowed. Authors must ensure that their Methods section includes adequate experimental and characterization data necessary for others in the field to reproduce their work.
\section{Experimental Results}\label{Sec:Experiment}
This section presents the experimental results of \framework. We first evaluate the effectiveness of the encryption component of \framework (i.e., \pricirc). Then, we show the importance of \framework to enable accurate inference on the encrypted data by the results on the noiseless quantum simulator. Finally, we evaluate \framework on the noisy quantum simulator and actual quantum computer (i.e., IBMQ Manila) to show the effect of quantum noise on \framework. 
% ~\todo{More results if necessary}

% We finally show the visualization results on some data samples encrypted by \pricirc to demonstrate that it can indeed effectively protect data.

\subsection{Experimental Setting}

\noindent\textbf{Dataset}. We evaluate~\framework on MNIST-2 (class), -3 (class); Fashion-2 (class), -3 (class). When evaluated on the noiseless simulator, the data are downsampled to a resolution of $4\times 4$ from $28\times 28$, which needs 4 data qubits with amplitude encoding. When evaluated in noisy environments (i.e., noisy simulator of IBMQ Manila and IBMQ Manila), the data is downsampled to a resolution of $4\times 2$ with 3 data qubits. In this case, due to the limited quantum resources, we use 100 samples and 150 samples from MNIST-2 and MNIST-3 for evaluation, respectively.

\noindent\textbf{Security Keys.} To generate the security keys, the range of random sampling of rotation angles is $[\frac{1\pi}{8}, \frac{7\pi}{8})$ and the permutation matrix $P$ is implemented with random generation of \CNOT gates. When evaluated on the noiseless simulator, for each encryption setting (i,e., number of secure qubits, dataset), we randomly generated 8 security keys. When evaluated in noisy environments, for each encryption setting, we randomly generated 3 security keys. 

\noindent\textbf{Metrics.} We use the peak signal-to-noise ratio (PSNR) to serve as a quantitative metric to evaluate the difference between two images. The smaller the PSNR is, the larger the difference is, indicating a better quality of encryption and security. Since amplitude encoding requires normalization of the features within the original data, it introduces differences between the quantum data and the original data. Therefore, we computed the PSNR between the original data and the quantum data (i.e., the PSNR when the number of secure qubits is 0) to be the baseline for the comparison of PSNR.

%uniformly denoted as K1-K8

% for each evaluated dataset, the encryption subcircuits are built using 1 and 2 secure qubits with \pricirc, separately

\noindent\textbf{QNNs.} All the evaluated child QNNs are implemented with TorchQuantum~\cite{hanruiwang2022quantumnas}. They are trained with batch size 64 for 20 epochs, using the Adam optimizer with a learning rate of 0.05. They are then compiled and executed by Qiskit. Besides, for \primod, there are 6 options for quantum neural designs at each layer, including 5 different quantum neuron designs from~\cite{sim2019expressibility} in the neuron design pool and one identity operation.

\subsection{PriCircuit Effectively Protects Data}\label{subsec:exp_pricirc}
% Please add the following required packages to your document preamble:
% \usepackage{multirow}
\begin{table}[!t]
\small
\centering
\caption{Evaluation of PriCircuit}
\label{tab:table-attack}
\setlength\tabcolsep{5.5 pt}
\begin{tabular}{ccccc}
\hline
Dataset & \begin{tabular}[c]{@{}c@{}}\# Secure\\ Qubits\end{tabular} & \begin{tabular}[c]{@{}c@{}}Model \\ Source\end{tabular} & Accuracy (\%) & PSNR(dB) \\ \hline
\multirow{4}{*}{\begin{tabular}[c]{@{}c@{}}MNIST-2 \end{tabular}} & \multirow{2}{*}{0} & User & 99.06 & \multirow{2}{*}{34.28} \\
 &  & Attacker & 99.06 &  \\
 & 1 & Attacker & 54.10 \textpm\ 16.06 & 19.35 \textpm\ 2.21 \\
 & 2 & Attacker & 52.23 \textpm\ 6.23 & 18.21 \textpm\ 8.37 \\ \hline
\multirow{4}{*}{\begin{tabular}[c]{@{}c@{}}MNIST-3 \end{tabular}} & \multirow{2}{*}{0} & User & 92.56 & \multirow{2}{*}{34.82} \\
 &  & Attacker & 92.29 &  \\
 & 1 & Attacker & 39.05 \textpm\ 6.12 & 20.00 \textpm\ 2.18 \\
 & 2 & Attacker & 36.29 \textpm\ 8.90 & 18.69 \textpm\ 8.31 \\ \hline
% \multirow{4}{*}{\begin{tabular}[c]{@{}c@{}}MNIST-4 \\ (6 qubits)\end{tabular}} & \multirow{2}{*}{0} & User & 93.76 & \multirow{2}{*}{24.46} \\
%  &  & Attacker & 93.44 &  \\
%  & 1 & Attacker & 33.95 \textpm\ 11.84 & 18.89 \textpm\ 1.22 \\
%  & 2 & Attacker & 29.76 \textpm\ 14.16 & 18.06 \textpm\ 0.25 \\ \hline
\multirow{4}{*}{\begin{tabular}[c]{@{}c@{}}Fashion-2 \end{tabular}} & \multirow{2}{*}{0} & User & 87.19 & \multirow{2}{*}{26.06} \\
 &  & Attacker & 86.54 &  \\
 & 1 & Attacker & 58.97 \textpm\ 17.33 & 18.15 \textpm\ 2.06 \\
 & 2 & Attacker & 54.45 \textpm\ 13.64 & 16.7 \textpm\ 0.50 \\ \hline
\multirow{4}{*}{\begin{tabular}[c]{@{}c@{}}Fashion-3 \end{tabular}} & \multirow{2}{*}{0} & User & 77.41 & \multirow{2}{*}{25.32} \\
 &  & Attacker & 77.38 &  \\
 & 1 & Attacker & 44.26 \textpm\ 10.86 & 17.81 \textpm\ 1.97 \\
 & 2 & Attacker & 35.82 \textpm\ 10.48 & 17.02 \textpm\ 1.47 \\ \hline
\end{tabular}
\end{table}

As Fig.~\ref{fig:qprivate} shows, we have two types of models for evaluation in this part, i.e., the model for user and the model for attacker. In general cases, the two models should target different types of tasks with the same dataset. But in this experiment, we trained two different models which both target the classification task for simplicity.

Table~\ref{tab:table-attack} reports the results for the evaluation of \pricirc. It clearly shows that without encryption, the data stolen from the user can be utilized by the attacker with its own model effectively. More specifically, the performance of the model for attacker is close to that of the model for user and thus verifies the threat model in Fig.~\ref{fig:qprivate}. For example, on MNIST-3, the accuracy of the model for attacker is 92.29 \% while that of the model for user is 92.56 \%.

When the data is encrypted by \pricirc, the security of data can be preserved. It means that the model for the attacker performs badly on the encrypted data while the difference between the encrypted data and the original data is increased significantly. For example, with only 1 secure qubit,~\pricirc can reduce the accuracy and PSNR to 39.05 \% and 20.0 dB on average on MNIST-3, respectively.

Moreover, increasing the number of secure qubits can further enhance the data security. For instance, on MNIST-3, the accuracy drops from 39.05 \% to 36.29 \% and the PSNR reduces from 20.0 dB to 18.69 dB respectively when increasing the secure qubits from 1 to 2.

\begin{figure*}[!t]
\centering
\includegraphics[width=\textwidth]{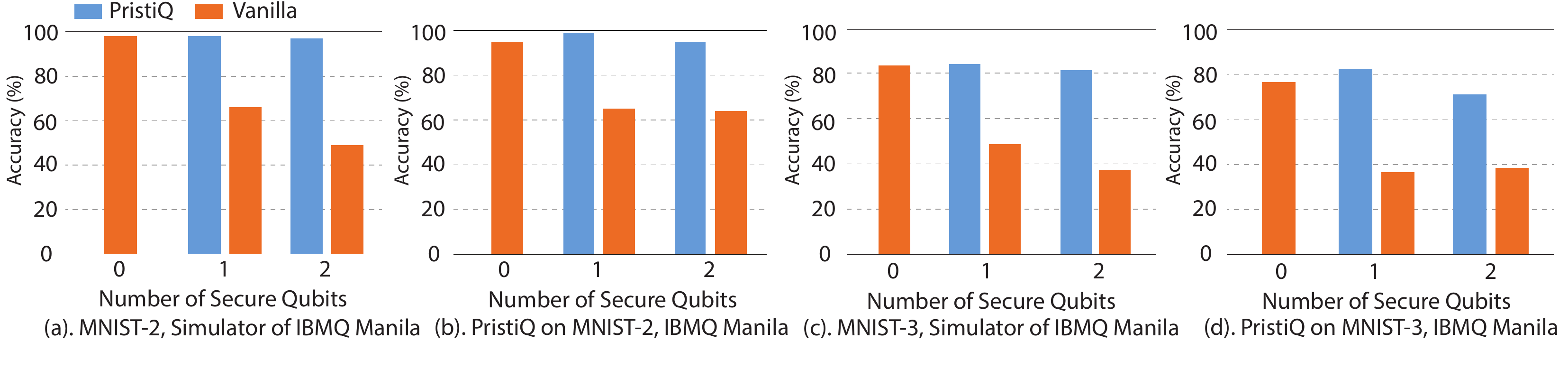}
\caption{Results of \framework for MNIST-2 on (a) simulator of IBMQ Manila and on (b) IBMQ Manila, and for MNIST-3 on (c) simulator of IBMQ Manila and on (d) IBMQ Manila.}
\label{fig:QC_Results}
\end{figure*}
\subsection{PristiQ Can Enable Accurate Inference While Protecting Data}\label{subsec:exp_pristiq_ideal}
% Please add the following required packages to your document preamble:
% \usepackage{multirow}
\begin{table}[t]
\small
\centering
\caption{Evaluation of PristiQ}
\label{tab:table-2}
\setlength\tabcolsep{2 pt}
\begin{tabular}{lccccc}
\hline
\multicolumn{1}{c}{Dataset} & \begin{tabular}[c]{@{}c@{}}\# Secure \\ Qubits\end{tabular} & \begin{tabular}[c]{@{}c@{}}Model \\ Type\end{tabular} & \# Param & \begin{tabular}[c]{@{}c@{}}Circuit \\ Length\end{tabular} & Accuracy (\%) \\ \hline
\multirow{5}{*}{\begin{tabular}[c]{@{}l@{}}MNIST-2 \end{tabular}} & 0 & User (Vanilia) & 56 & 82 & 99.06 \\
 & \multirow{2}{*}{1} & User (Vanilia) & 56 & 82 & 53.65   \textpm\ 14.98 \\
 &  & User (PristiQ) & \textbf{59.5} & \textbf{96.88} & \textbf{99.15 \textpm\ 0.05} \\
 & \multirow{2}{*}{2} & User (Vanilia) & 56 & 82 & 49.81 \textpm\ 4.79 \\
 &  & User (PristiQ) & \textbf{65} & \textbf{135} & \textbf{99.10 \textpm\ 0.04} \\ \hline
\multirow{5}{*}{\begin{tabular}[c]{@{}l@{}}MNIST-3 \end{tabular}} & 0 & User (Vanilia) & 44 & 52 & 92.56 \\
 & \multirow{2}{*}{1} & User (Vanilia) & 44 & 52 & 40.07 \textpm\ 9.52 \\
 &  & User (PristiQ) & \textbf{48.5} & \textbf{72.25} & \textbf{95.08 \textpm\ 0.27} \\
 & \multirow{2}{*}{2} & User (Vanilia) & 44 & 52 & 40.22 \textpm\ 9.96 \\
 &  & User (PristiQ) & \textbf{58.5} & \textbf{115.25} & \textbf{95.03 \textpm\ 0.24} \\ \hline
% \multirow{5}{*}{\begin{tabular}[c]{@{}l@{}}MNIST-4 \\ (6 qubits)\end{tabular}} & 0 & User (Vanilia) & 174 & 242 & 93.76 \\
%  & \multirow{2}{*}{1} & User (Vanilia) & 174 & 242 & 33.44   \textpm\ 11.07 \\
%  &  & User (PristiQ) & \textbf{141.75} & \textbf{246.63} & \textbf{95.25 \textpm\ 0.24} \\
%  & \multirow{2}{*}{2} & User (Vanilia) & 174 & 242 & 32.02 \textpm\ 8.81 \\
%  &  & User (PristiQ) & \textbf{151.5} & \textbf{332.63} & \textbf{95.17 \textpm\ 0.13} \\ \hline
\multirow{5}{*}{\begin{tabular}[c]{@{}l@{}}Fashion-2 \end{tabular}} & 0 & User (Vanilia) & 80 & 103 & 87.19 \\
 & \multirow{2}{*}{1} & User (Vanilia) & 80 & 103 & 58.97   \textpm\ 17.35 \\
 &  & User (PristiQ) & \textbf{41.5} & \textbf{59.38} & \textbf{88.30 \textpm\ 0.19} \\
 & \multirow{2}{*}{2} & User (Vanilia) & 80 & 103 & 54.39 \textpm\ 13.54 \\
 &  & User (PristiQ) & \textbf{45.5} & \textbf{79.5} & \textbf{88.23 \textpm\ 0.16} \\ \hline
\multirow{5}{*}{\begin{tabular}[c]{@{}l@{}}Fashion-3 \end{tabular}} & 0 & User (Vanilia) & 48 & 46 & 77.41 \\
 & \multirow{2}{*}{1} & User (Vanilia) & 48 & 46 & 36.97   \textpm\ 9.26 \\
 &  & User (PristiQ) & \textbf{57.5} & \textbf{90.38} & \textbf{79.13 \textpm\ 0.09} \\
 & \multirow{2}{*}{2} & User (Vanilia) & 48 & 46 & 36.80 \textpm\ 13.42 \\
 &  & User (PristiQ) & \textbf{67} & \textbf{135.88} & \textbf{79.07 \textpm\ 0.10} \\ \hline
\end{tabular}
\end{table}

Table~\ref{tab:table-2} shows the results for the evaluation of \framework on the noiseless simulator, where the model ``User (Vanilla)'' refers to the model for user optimized on the original data. Without encryption (i.e., the number of secure qubits is 0),  it is clearly shown that only using \pricirc to protect data will lead to the performance collapse of the model for user. For example, on MNIST-2, the accuracy drops from 99.06 $\%$ to 49.81 $\%$ on average with 2 secure qubits. 

By applying \framework to building the model, where \pricirc is included, the performance of the model (i.e., ``User (PristiQ)'' in Table~\ref{tab:table-2}) can be recovered on all the evaluated datasets. For example, on MNIST-2, the accuracy is recovered to 99.10\% on average when encrypted with 2 secure qubits, which is even improved by 0.04 $\%$ compared with the model optimized on the original data. 

% For the slight improvement, there are several reasons. Firstly, more learnable parameters are introduced, which indicates a larger model capacity. Secondly, the circuit length for the model is increased, which means more complicated operations are applied to process the input. Moreover, the number of features is also increased after encryption, which has the potential to be learned better if the model is customized to the encrypted features. Note that due to the limitation of space, we remove the standard deviation for the number of parameters and circuit length in Table~\ref{tab:table-2}.

% ~\todo{enumerate the data points in the table??}

% ~\todo{at this point they may not know we re-design the QNN architecture ?}

% \todo{Do we need to explain the reason here - directly train on the encrypted training data}

% \todo{Highlight baseline is NAS generated? For the noisy part, not all of them is OK}

\subsection{Evaluation of PristiQ in Noisy Environments}\label{subsec:exp_pristiq_noise}

Fig.~\ref{fig:QC_Results} shows the effect of quantum noise on the performance of \framework on MNIST-2 and MNIST-3. In Fig.~\ref{fig:QC_Results}, the orange bar (Vanilla) denotes the model for user optimized on the original data while the blue bar (PristiQ) denotes the model optimized on the encrypted data with \framework. Fig.~\ref{fig:QC_Results} (a) and (c) show the results on the noisy simulator of IBMQ Manila, while Fig.~\ref{fig:QC_Results} (b) and (d) show the results on the actual quantum computer (i.e., IBMQ Manila).

% ~\todo{do we need to explain the setting here?}~\todo{Claim we pick the best result? run 3 security keys?}

% Compared with the results on the noiseless simulator in Table~\ref{tab:table-2}, it is obvious that the noise can indeed affect the performance of QNN model. For example, (Vanilla).~\todo{Do we need this part? examples for MNIST-2? Test set is not consistent.}
For the vanilla model, we can conclude that its performance degrades significantly on the encrypted data generated by \pricirc in noisy environments. For example, with 2 secure qubits, the accuracy decreases from 98 \% to 49 \% on the noisy simulator and drops from 95 \% to 64 \% on the actual quantum computer on MNIST-2.

By applying \framework to optimizing the model, the performance can still be recovered even with the quantum noise. For instance, with 2 secure qubits, the accuracy can be recovered from 49 \% to 97 \% on the noisy simulator and increased from 64 \% to 95 \% on the actual quantum computer on MNIST-2.

%---------------------------------------------
% results in appendix

% \input{text/add_exp}
%\vspace{-6pt}
\section{Conclusion}\label{Sec:Conclusions}
In this paper, we made the very first exploration of the problem of data security for QML in the cloud. We proposed PristiQ, a framework to preserve data security in QML. By creating an encryption subcircuit with a user-defined security key, the important information in the original data is protected. Besides, PristiQ utilizes an automatic model optimizer to achieve high performance on the encrypted data. Extensive experiments in the noiseless quantum simulator and noisy quantum environments are conducted to show the effectiveness of PristiQ. Moreover, the design philosophy of PristiQ, a cross-layer co-design from the circuit level and compiler level to the application level, could be applied to guide the design of a secure quantum computing system in applications beyond QML.

\section*{Acknowledgment}
This work is partly supported by the National Science Foundation (NSF) OAC-2311949 and OAC-2320957.
The research used IBM Quantum resources via the Oak Ridge Leadership Computing Facility at the Oak Ridge National Lab, which is supported by the Office of Science of the U.S. Department of Energy under Contract No. DE-AC05-00OR22725. This project was also supported by resources provided by the Office of Research Computing at George Mason University (URL: https://orc.gmu.edu) and funded in part by grants from the National Science Foundation (Award Number 2018631). And we also thank Mason's QSEC and C-TASC centers for their support.

\bibliographystyle{IEEEtranS}
\bibliography{quantum, reference, new_reference}

\end{document}